\documentclass[prl,aps,twocolumn,showpacs,raggedbottom,eprint,superscriptaddress,10pt,longbibliography]{revtex4-2} 

\usepackage{graphicx}

\usepackage{xcolor}

\usepackage{float}

\usepackage[utf8]{inputenc}

\usepackage{amsmath}
\usepackage{braket}
\usepackage{dsfont}

\usepackage{xspace}

\let\vec\mathbf

\usepackage{hyperref}
\hypersetup{
	colorlinks=true,
	linkcolor=blue,
	filecolor=blue,
	citecolor = black,      
	urlcolor=cyan,
}

\newcommand{\blst}{\begin{lstlisting}}
\newcommand{\est}{\end{lstlisting}}

\newcommand{\bit}{ \begin{itemize}{}  }
\newcommand{\eit}{\end{itemize}}

\newcommand{\be}{\begin{enumerate} \itemsep -4pt  }
\newcommand{\ee}{\end{enumerate}}

\newcommand{\bi}{\begin{itemize}   } %
\newcommand{\ei}{\end{itemize}}

\newcommand{\bma}{\begin{math}} 
\newcommand{\ema}{\end{math}}

\newcommand{\bc}{\begin{columns}} 
\newcommand{\ec}{\end{columns}}

\newcommand{\bbl}{\begin{block}} 
\newcommand{\ebl}{\end{block}}

\newcommand{\bflsh}{\begin{flashcard}}
\newcommand{\eflsh}{\end{flashcard}}

\newcommand{\bfl}[2]{\begin{flashcard}{#1} {#2} \eflsh}

\newcommand{\beq}{\begin{equation} \vspace{-0em}} 
\newcommand{\eeq}{\vspace{-0.em} \end{equation}}

\newcommand{\beqs}{\begin{equation*}}
\newcommand{\eeqs}{\end{equation*}}

\def\beqa{\begin{eqnarray}}
\def\eeqa{\end{eqnarray}}

\newcommand{\beal}{\begin{align}}
\newcommand{\eeal}{\end{align}}

\newcommand{\bvr}{\begin{verbatim}}
\newcommand{\evr}{\end{verbatim}}

\newcommand{\unit}[1]{\ensuremath{\, \mathrm{#1}}}                     		%

\newcommand{\remove}[1]{}

\usepackage{catchfilebetweentags}
\newif\ifsupp
\supptrue

\begin{document}
	
\title {Topological bands in the continuum using Rydberg states}

\author{Sebastian~Weber}
\thanks{These authors contributed equally.}
\affiliation{Institute for Theoretical Physics III and Center for Integrated Quantum Science and Technology, University of Stuttgart, 70550 Stuttgart, Germany}
\author{Przemyslaw~Bienias}
\thanks{These authors contributed equally.}
\affiliation{Joint Quantum Institute and Joint Center for Quantum Information and Computer Science, NIST/University of Maryland, College Park, MD, 20742, USA}
\author{Hans~Peter~B\"{u}chler}
\affiliation{Institute for Theoretical Physics III and Center for Integrated Quantum Science and Technology, University of Stuttgart, 70550 Stuttgart, Germany}

\date{\today}

\begin{abstract}
	The quest to realize topological band structures in artificial matter is strongly focused on lattice systems, and only quantum Hall physics is 
	known to appear naturally also in the continuum. In this letter, we present a proposal  based on a two-dimensional cloud of atoms dressed to Rydberg states, 
	where excitations propagate by dipolar exchange interaction, while the Rydberg blockade phenomenon naturally gives rise to a characteristic length scale, suppressing the hopping on short distances. Then, the system becomes independent of the atoms' spatial arrangement and can be described by a continuum model.  We demonstrate the appearance of a topological band structure in the continuum characterized by a Chern number $C=2$ and show that edge states appear at interfaces tunable by the atomic density.
\end{abstract}


\maketitle

Band structures characterized by topological invariants are at the heart of a plethora of phenomena robust to local perturbations because the topological invariants can only change in case of band closings \cite{Hasan2010, Qi2011}. In finite systems or at interfaces,  topologically protected edge states emerge, whose robustness makes them promising for coherent transport of quantum information \cite{Halperin1982, Dlaska2017, Lang2017}. In combination with strong interactions, topological bands are candidates for realizing topologically ordered states exhibiting anyonic excitations \cite{Nayak2008}. Remarkably, such phenomena occur in both continuum and lattice models. Especially, the renowned integer \cite{Klitzing1980} and fractional \cite{Tsui1982} quantum Hall effect arise naturally in continuous two-dimensional electron systems  subject to strong magnetic fields. 
In contrast, the efforts to realize a wide variety of topological band structures and topological states of matter mainly focus on lattice models. In this letter, we bridge this gap and demonstrate a topological band structure characterized by a Chern number $C=2$ in the continuum based on the intrinsic spin-orbit coupling with dipolar exchange interaction.

The quest to realize topological band structure in artificial matter was pioneered by the theoretical proposal \cite{Jaksch2003} to 
realize the Hofstadter butterfly \cite{Hofstadter1976, Thouless1982} with cold atomic gases in optical lattices.
Since then, several promising experimental platforms have emerged such as classical phononic setups 
\cite{Suesstrunk2015, SerraGarcia2018}, and photonic nanostructures \cite{Hafezi2013, Lu2016, Milicevic2017}, solid-state devices 
\cite{Tsukazaki2007, Koenig2017, Drozdov2014}, and cold atomic gases \cite{Jotzu2014, Aidelsburger2015, Stuhl2015, Mancini2015, Flaeschner2016,Chalopin2020}.
While all these platforms work on lattices, fewer experiments pursue the challenge to achieve topological states in the continuum, relying on implementing the physics of magnetic fields either by artificial magnetic fields \cite{Lin2009,Dalibard2011,Goldman2014} of fast rotating traps \cite{Fletcher2019}. Rydberg atoms and polar molecules have
recently emerged as a complementary platform to generate topological band structures, using the dipolar exchange interaction \cite{Peter2015, Weber2018, Yao2013a}. A fundamental advantage of this approach is that the time-reversal symmetry is broken by a homogeneous magnetic field \cite{Weber2018}, therefore avoiding potentially problematic heating present in approaches based on Floquet engineering and experimentally challenging strong spatially-inhomogeneous light fields. The underlying principle has recently been experimentally verified using Rydberg 
atoms  \cite{Lienhard2020} and the potential of Rydberg platforms  to realize topological states of matter 
has been demonstrated by observing a symmetry-protected topological phase \cite{deLeseleuc_2019}.

\begin{figure}[htb]
	\includegraphics{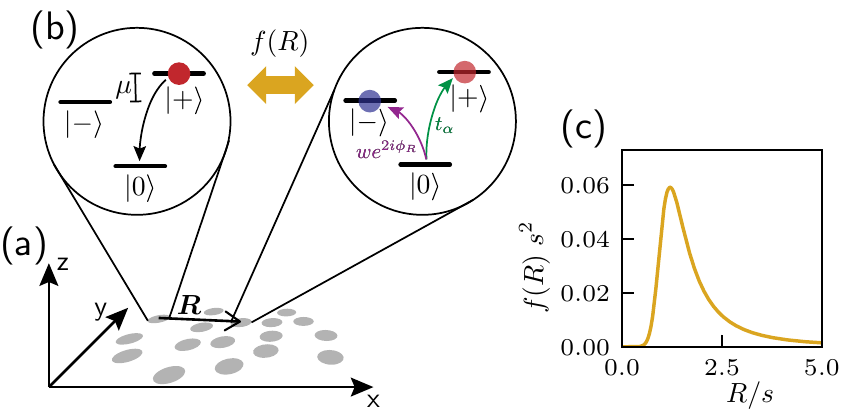}
	\caption{
		Setup. (a) Two-dimensional homogeneous cloud of atoms with quantization axis $z$. (b) Each atom can be either in the ground  state $\ket{0}$ with 
		the magnetic quantum number $m_0$ or a dipole-coupled excited state $\ket{+}$ or $\ket{-}$, having $m_\pm=m_0 \pm 1$ as an internal degree of freedom. 
		The energy difference between these states is $\mu = E_+-  E_-$. We consider the case of a single excitation in the cloud of atoms. The excitation can hop 
		from one atom to another by dipolar exchange interaction that keeps the excitation's internal degree of freedom constant or changes it. In the latter case, the 
		excitation picks up a non-trivial phase (spin-orbit coupling). (c) We design the interaction potentials to have the distance dependence given by the function 
		$f(R)$ with short-distance cut-off $s$, chosen to be much larger than the average distance between the atoms.}
	\label{fig:setup}
\end{figure}

Here, we show that the intrinsic spin-orbit coupling by the dipolar exchange interaction of Rydberg atoms can lead to topological bands in the continuum.  
The basic setup consists of cold atoms in the frozen regime, dressed to Rydberg states; a system
studied
extensively  in the past \cite{Balewski2014, Zeiher2016, Macri2014, Honer2010, Glaetzle2015, Jau2016, VanBijnen2015, Gaul2016,Grass2018,Belyansky2020d,Belyansky2019a,Guardado-Sanchez2020,Young2020}.  
The combination of two ideas underlies our proposal: (i) The possibility to engineer non-trivial topological band structures using the intrinsic spin-orbit coupling of dipolar exchange interaction \cite{Weber2018,Lienhard2020}, and (ii) the Rydberg blockade phenomenon, which suppresses the presence of two Rydberg excitation on short distances \cite{Lukin2001}, and introduces a natural length scale by the blockade radius, see Fig.~\ref{fig:setup}. Then, the admixture of Rydberg states by the dressing lasers allows for the hopping of excitations by the dipolar exchange interaction. Still, this hopping becomes quenched on short distances due to the Rydberg blockade phenomenon.  Remarkably, the short distance cut-off by the Rydberg blockade renders the system independent of the precise atomic distribution of the atoms.  We find that the emerging band structure in the topological regime is characterized by the Chern number $C=2$, and the system 
exhibits edge states at boundaries or interfaces, which can be tuned by the atomic density. 

We start with the description of our systems. A cloud of atoms is confined to two dimensions with the quantization axis and a magnetic field perpendicular to the plane, see Fig.~\ref{fig:setup}.
Each atom possesses a V-level structure formed by the ground state $\ket{0}$ with magnetic quantum number $m_0$ and excited states $\ket{+}$ and $\ket{-}$. The excited states differ by their magnetic quantum number $m_{\pm} = m_0 \pm 1$, constituting an internal degree of freedom. The states of the V-level structure are dressed with Rydberg states, giving rise to an effective dipolar exchange interaction, leading to a single excitation hopping at large interatomic distances $R$ in analogy to studies on a lattice ~\cite{Peter2015, Weber2018, Yao2013a}. However, in our case,
the conventional Rydberg blockade suppresses the Rydberg states' admixture on distances shorter than a characteristic length scale $s$. It therefore quenches the dipolar exchange interaction and the hopping of excitations.  In the following, we focus on the dynamics of such a single excitation.  If the distance between the atoms in the cloud is much smaller than the cut-off distance, i.e. $ns^2 \gg 1 $ with the homogeneous density $n$ of atoms in the ground state, we can express the excitations in terms of field operators $\alpha^{\dagger}(\vec{r})$ and $\beta^{\dagger}(\vec{r})$ ($\alpha(\vec{r})$ and $\beta(\vec{r})$), which create (annihilate) a $\ket{+}$ and $\ket{-}$ excitation at the position $\vec{r}$, respectively.
Then, the Hamiltonian describing the dynamics of excitations under the dipolar exchange interaction takes the form
\begin{align}
H= & n \iint d{\bf r} d{\bf r}'
\psi^\dagger(\vec{r}) \begin{pmatrix} -T_{\alpha}(R) & W(R) e^{-2i\phi_{\vec{R}}}  \\ W(R) e^{2i\phi_{\vec{R}}}  & -T_{\beta}(R) \end{pmatrix}
\psi(\vec{r}') \nonumber
\\
&+ \int d{\bf r} \: 
\psi^\dagger(\vec{r}) 
\begin{pmatrix} \mu/2 & 0 \\ 0 & - \mu/2 \end{pmatrix}
\psi(\vec{r}) \label{eqn:dipolar}.
\end{align}
Here, $\psi^\dagger(\vec{r}) = \begin{pmatrix} \alpha^\dagger(\vec{r}), & \beta^\dagger(\vec{r}) \end{pmatrix}$ is the spinor field operator,  while  $\vec{R} = \vec{r}'-\vec{r}$ is the relative distance and $R = |\vec{r}'-\vec{r}|$. The last term takes into account the energy difference between the two excitations $\mu = E_{+} -  E_{-}$, tunable by the strength of the magnetic field applied along the quantization axis.
The first term in Eq.~(\ref{eqn:dipolar}) describes two different hopping processes: First, the $\ket{+}$ and $\ket{-}$ excitations can hop mediated by the amplitudes $T_{\alpha, \beta}(R)$ while conserving the internal angular momentum. Second, the excitations can change their internal angular momentum by a hopping process with amplitude $W(R)$, i.e., a  $\ket{\pm}$ excitation becomes a $\ket{\mp}$.  The phase factor accompanies the change of internal angular momentum of the excitation $e^{-2i\phi_{\vec{R}}}$, where $\phi_{\vec{R}}$ is the polar angle of $\vec{R}$. Note that for an excitation hopping on a closed loop, the total collected phase is independent of the chosen coordinate system and cannot be gauged away. This phase accounts for a spin-orbit coupling and guarantees that the total angular momentum is conserved, i.e., the change in internal angular momentum of the excitation is transferred into orbital angular momentum.

\begin{figure}
	\includegraphics{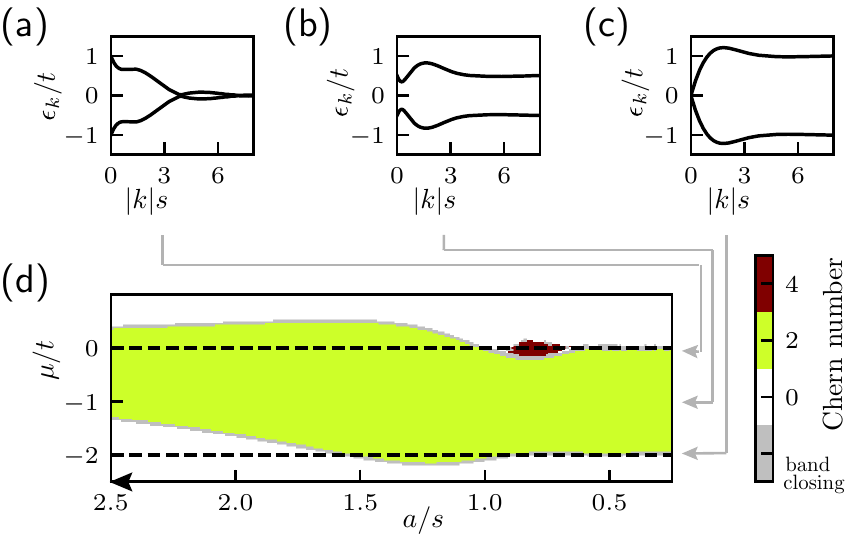}
	\caption{ (a-c) Isotropic band structure of the continuum model for different $\mu/t$ calculated for an exemplary value of $w / t = 3$ and  $\overline{t}=0$. 
		Between the band closings at $\mu/t = -2$ and $\mu/t = 0$, the Chern number is $C=2$. (d) The phase diagram of the dicretized model 
		on a square lattice approaches the continuum model (dashed lines) when the lattice constant $a$ gets smaller than the cut-off distance $s$.}
	\label{fig:transition}
\end{figure}

The effective hopping Hamiltonian~(\ref{eqn:dipolar}) gives rise to a band structure with two bands. In the following, we focus on characterizing the topological properties of this band structure. A discussion on optimal parameters for an experimental realization and the detailed microscopic interaction potentials is presented at the end. As the precise shape of the hopping amplitudes is not essential for the topological properties of the band structure, we assume that each hopping amplitude obeys the same distance dependency given by the cut-off function
\begin{equation}
f(R)
= \frac{9s}{5\pi^2}\;\frac{R^9}{(R^6+s^6)^2}\;.
\label{eqn:cutoff}
\end{equation}
with the characteristic cut-off distance $s$. The normalization is chosen such that $\int \text{d}^2 R\; f(R)=1$. We write $T_{\alpha}(R)\;n = t_\alpha f(R)$ with 
the interaction energy  $t_\alpha=\int \text{d}^2 R\; T_{\alpha}(R)\;n$. Analogously, we introduce the interaction energies $t_\beta$ and $w$. By Fourier transforming 
the field operators as $\psi_{\vec{k}}^\dagger = \int d{\bf r} e^{i\vec{k}\vec{r}} \psi^\dagger(\vec{r})/\sqrt{V}$ with quantization volume $V$, we obtain 
the Hamiltonian in momentum space
\begin{equation}
H = \sum_{\vec{k}} \psi_{\vec{k}}^\dagger  \left(
\overline{t} \epsilon^0_{\vec{k}} \mathds{1} + \vec{n}(\vec{k}) \cdot  \boldsymbol{\sigma}
\right) \psi_{\vec{k}}\;,
\end{equation}
Here, $\boldsymbol{\sigma}$ is the vector of Pauli matrices, and we have introduced the vector
\begin{equation}
\vec{n}(\vec{k}) = \begin{pmatrix} 
w  \left(  \epsilon^2(\vec{k})+ \epsilon^{-2}(\vec{k})\right) /2\\
w \left( \epsilon^2(\vec{k})-  \epsilon^{-2}(\vec{k}) \right)/2 i\\
\mu / 2 + t \epsilon^0_{\vec{k}}
\end{pmatrix}
\end{equation}
with the average hopping strength $\overline{t} = (t_\beta+t_\alpha)/2$ and $t=(t_\beta-t_\alpha)/2$. 
Furthermore, we have made use of the Fourier transformation for the hopping amplitudes
\begin{equation}
\epsilon^m_{\vec{k}} = \int d{\bf  r}\; e^{i\vec{k}\vec{r}+im\phi_{\vec{r}}}f(|\vec{r}|) = 2\pi (-1)^m \text{e}^{\text{i}m \phi_{\vec{k}}} F_m(|\vec{k}|)\;,
\label{eqn:continuum_ft}
\end{equation}
where $F_m$ is the Hankel transform of order $m$ of the cut-off function~(\ref{eqn:cutoff}). In our continuum model, momenta can have arbitrary high values in contrast to lattice models. However, interesting physics happen only at momenta comparable to the characteristic momentum $2 \pi/s$ defined by the cut-off distance $s$. As the Hamiltonian is block-diagonal in momentum space with block-size $2\times2$, the band structure consists of two bands whose Chern number can be calculated as a winding number of the three-dimensional unit vector $\hat{\vec{n}}(\vec{k}) = \vec{n}(\vec{k}) / |\vec{n}(\vec{k})|$  \cite{Peter2015}. Note, that $\hat{\vec{n}}(\vec{k})$ is pointing into the same direction $\hat{\vec{z}}\;\text{sign}(\mu)$ for all $|\vec{k}| \rightarrow \infty$, and therefore we can compactify the infinite Brillouin zone to the sphere $S^2$. We obtain for an infinite quantization volume the Chern number for the lower band as
\begin{equation}
C = \frac{1}{4\pi} \int d{\bf  k}\;\left( \partial_{\vec{k}_x} \hat{\vec{n}}(\vec{k}) \times \partial_{\vec{k}_y} \hat{\vec{n}}(\vec{k})  \right) \cdot \hat{\vec{n}}(\vec{k})\;.
\end{equation}
For $- 2 < \mu/t < 0$, the system is in its topological phase with Chern number $C=2$, see Fig.~\ref{fig:transition}(b). This condition implies that  $\mu \neq 0$ and $t \neq 0$.
Outside this parameter regime, the system is in the topologically trivial phase. At the phase boundaries, the bandgap closes, see Fig.~\ref{fig:transition}(a,c). Note that these 
findings are independent of the value of $\overline{t}$ and $w$ as long as $w \neq 0$. However, the bandstructure has a quantitative dependence on these parameters. Finally, 
we point out that $\mu \neq 0$ or $t \neq 0$ give rise to a broken time-reversal symmetry as required for the appearance of a non-zero Chern number \cite{Ryu2010}.

\begin{figure}
	\includegraphics{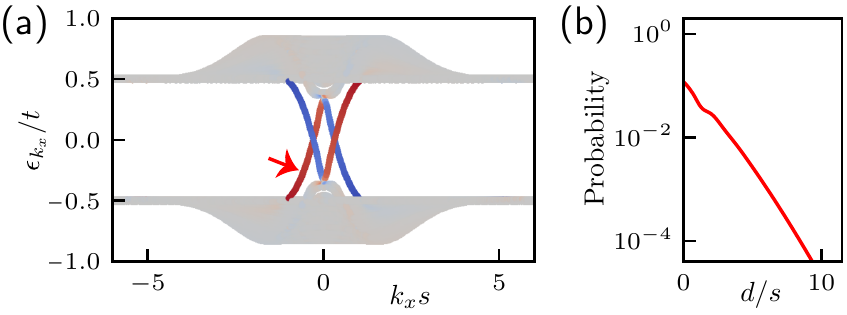}
	\caption{ (a) For a semi-infinite cloud of atoms, the band structure in the topological sector hosts two edge states per boundary, bridging the bandgap. 
		The two edge states colored in red (blue) are localized at the cloud's upper (lower) boundary. The numerical analysis is based on the discretized model 
		with lattice spacing $a/s =1/6$ and width $l = 20 s$
		(b) The probability of finding an excitation in an edge state decreases exponentially with the distance $d/s$ from the edge, as shown exemplarily for the mode 
		marked with the red arrow.}
	\label{fig:edgestates}
\end{figure}

An interesting aspect is to connect the continuum model to a lattice model by discretizing the Hamiltonian on a square lattice with a lattice constant $a$. This 
discretized model is suitable to study numerically the appearance of edge states, and for large lattice spacing $a\gg s$, we recover the models studied previously 
\cite{Peter2015,Weber2018}. We introduce the operators $a^{\dagger}_i$ and $b^{\dagger}_i$ ($a_i$ and $b_i$), which create (annihilate) a $\ket{+}$ or $\ket{-}$ 
excitation at the lattice site $i$, respectively. With $\vec{R}_{ij} = \vec{r}_j - \vec{r}_i$ being the distance vector pointing from site $i$ to $j$, the Hamiltonian reads
\begin{align}
H_\text{sq} &= 
\sum_{i,j}
a^2 f(|\vec{R}_{ij}|)
\begin{pmatrix} a_i \\ b_i \end{pmatrix}^\dagger
\begin{pmatrix} -t_\alpha & w e^{-2i\phi_{\vec{R}_{ij}}} \\ w e^{2i\phi_{\vec{R}_{ij}}} & -t_\beta \end{pmatrix}
\begin{pmatrix} a_j \\ b_j \end{pmatrix} \nonumber \\
&+
\sum_{i}
\begin{pmatrix} a_i \\ b_i \end{pmatrix}^\dagger
\begin{pmatrix} \mu/2 & 0 \\ 0 & - \mu/2 \end{pmatrix}
\begin{pmatrix} a_i \\ b_i \end{pmatrix}\;.
\label{eqn:discrete}
\end{align}
It approaches the Hamiltonian of the continuum model (\ref{eqn:dipolar}) in the limit $a/s \rightarrow 0$. In momentum space, the periodic Brillouin zone of the square lattice 
becomes the infinite Brillouin zone of the continuum model, and the discrete Fourier transform of the interaction 
$\epsilon_{\text{sq},\vec{k}}^m = \sum_{i} a^2 e^{i\vec{k}\vec{r}_i+im\phi_{\vec{r}_i}}f(|\vec{r}_i|)$ becomes the continuous Fourier transform (\ref{eqn:continuum_ft}). The 
convergence is also visible in the phase diagram Fig.~\ref{fig:transition}(d), where we calculated the Chern number for different values of $\mu/t$ as a function of $a/s$ \cite{Fukui2005}. 
Already for a lattice constant $a$, which is half of the cut-off distance $s$, we obtain the phase boundaries of the continuum model in good approximation. 
For such a lattice spacing, the nearest neighbor interaction is strongly suppressed due to the fast decay of the cut-off function (\ref{eqn:cutoff}) on short distances. 
Interestingly, for $a/s \approx 0.8$, there is an additional phase with Chern number $C=4$.

The discretized version of the continuum model allows us to numerically calculate the band structure in a semi-infinite cloud of atoms, which is infinite in the $x$-direction and finite in the $y$-direction, see Fig.~\ref{fig:edgestates}. For the calculation, we have chosen a small lattice constant compared to the cut-off distance, $a/s=1/6$, to ensure convergence towards the continuum model. According to the bulk-boundary correspondence \cite{Essin2011}, the system hosts two edge states on each of the two boundaries of the semi-infinite 
cloud in the topological sector. We have chosen the semi-infinite cloud's width $l/s=20$, and indeed find two localized edge states within the band gap on each edge. 
We find that the edge states are exponentially localized at the edge for the studied system sizes; due to the slow decay of the dipolar exchange interaction, one can not exclude a transition to a power-law behavior at larger distances \cite{Vodola2016,Jager2020}.
Note that within the continuum model, the hopping strength $t \sim n$ depends linearly on the atomic density.  Therefore, the density in an inhomogeneous cloud can exhibit edge states as the ratio $\mu/t\propto 1/n$ 
becomes position-dependent with a topologically trivial phase in areas with low density and a topological phase in high-density regions. 

\begin{figure}
	\includegraphics{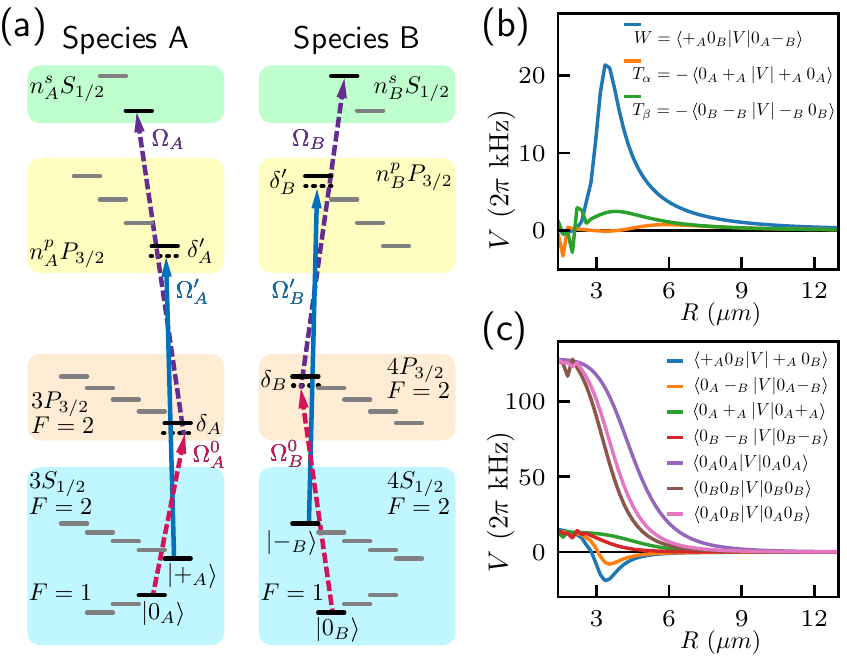}
	\caption{Experimental realization of the interaction potentials. (a)  The hyperfine states $\ket{0_A}$ and $\ket{+_A}$ of species~A~($^{23}$Na) are 
		dressed by Rydberg states, likewise $\ket{0_B}$ and $\ket{-_B}$ of species~B~($^{39}$K). The arrows visualize applied laser fields with detunings $\delta$ 
		and Rabi frequencies $2\Omega$. Dashed arrows symbolize EIT dressing, while solid arrows depict single-photon dressing. A static magnetic field $B_z$ 
		perpendicular to the cloud of atoms isolates the Rydberg states and brings the Rydberg pair states, with which the states $\ket{0_A-_B}$ and $\ket{+_A0_B}$ 
		are dressed, energetically close. (b,c) Effective exchange (b) and static (c) interaction between different pairs of the dressed hyperfine states for the following 
		parameters: $n_A^s = 61$,  $n_A^p = 60$, $n_B^s = 61$,  $n_B^p = 61$, $\Omega^0_{A,B}/(2\pi) = 7\unit{MHz}$, $\Omega_{A,B}/(2\pi)  = 35\unit{MHz}$, 
		$\Omega'_{A,B}/(2\pi) = 7\unit{MHz}$, $\delta_{A,B}/(2\pi)  =50 \unit{MHz}$, $\delta'_{A,B}/(2\pi)  =130 \unit{MHz}$, $B_z=-223\unit{G}$.}
	\label{fig:realization}
\end{figure}

Next, we discuss possible experimental realizations of the continuum model using Rydberg dressing.
We must use Rydberg states for the dressing, which are energetically isolated from other Rydberg states, especially those within the same Zeeman manifold. We apply a strong static magnetic field perpendicular to the atomic cloud to achieve the isolation, also for interatomic distances smaller than the cut-off distance. On the other hand, the  Rydberg states used for dressing the states $\ket{0+}$ and $\ket{-0}$ should be of similar energy to obtain a strong, effective interaction $W(R)$. 
In principle, a static electric field can compensate for the Zeeman splitting and shift the Rydberg pair states close to resonance, as proposed in Ref.~\cite{Weber2018}. 
However, if the Zeeman splitting is too large, this cannot be achieved without the electric field heavily mixing the Rydberg states. Thus, we propose a different approach. 
We suggest to realize the model with a homogeneous atomic cloud composed of two different species rather than a single species as in Ref.~\cite{Weber2018}, optically 
pumped into the hyperfine states $\ket{0_A}$ and $\ket{0_B}$ of species A and B. The excitations $\ket{+}$ and $\ket{-}$  correspond to the hyperfine states 
$\ket{+_A}$ and $\ket{-_B}$, respectively. Rydberg states weekly dress these four hyperfine states, and we use $^{23}$Na as species A and $^{39}$K as 
species B. We dress $\ket{0_A}$ with $\ket{s_A} =  \ket{n_A^s S_{1/2}, m_j = 1/2}$, $\ket{+_A}$ with $\ket{p_A} = \ket{n_A^p P_{3/2}, m_j = 3/2}$, $\ket{0_B}$ 
with $\ket{s_B} = \ket{n_B^s S_{1/2}, m_j = -1/2}$, and $\ket{-_B}$ with $\ket{p_B} =  \ket{n_B^p P_{3/2}, m_j = -3/2}$. For details on the dressing and suitable 
parameters, see Fig.~\ref{fig:realization}. The magnetic field applied to isolate the states is tuned, such that $\ket{s_A p_B}$ and $\ket{p_A s_B}$ are energetically 
close. We calculate the effective interaction by restricting the two-atom Hamiltonian to the manifold of the chosen hyperfine states within perturbation theory via 
the Schrieffer-Wolff transformation \cite{Schrieffer1966, Bravyi2011, Weber2017}, generalizing the numerical method in Ref.~\cite{Zeiher2016}. The hopping amplitudes
are shown in Fig.~\ref{fig:realization}(b) for optimal parameters. Furthermore, we show in the supplement \cite{supplementContinuum}, that the cut-off function in Eq.~(\ref{eqn:cutoff}) can be obtained analytically in fourth-order perturbation theory in case of energetically well separated Rydberg states.	
In addition to the  effective dipolar exchange interaction (\ref{eqn:cutoff}), such an experimental realization also yields static interactions, 
see Fig.~\ref{fig:realization}(c). The latter leads to a density-dependent shift of the chemical potential $\mu$. For a homogeneous system, this shift is everywhere in the bulk the same. By tuning the frequencies of the dressing lasers or the relative energy using the magnetic field, we can compensate for it and ensure that we are in the topological regime. At short interatomic distances, the Rydberg interaction unavoidably shifts some Rydberg states into resonance with the dressing lasers. However, the resulting divergencies in the potential curves are very sharp as they appear on short interatomic distances so that they might be ignored, particularly in the presence of dissipation (leading effectively to the resonance suppression), or could be avoided altogether by placing the atoms in an optical lattice with carefully chosen lattice spacing \cite{Derevianko2015a}. 
Besides, the level structure might be tuned further by applying an electric field. 

In summary, we demonstrated that the combination of broken time-reversal symmetry and spin-orbit coupling in the spatial continuum can give rise to topological bands. Our continuum model features topological bands with Chern number $C=2$ and exponentially localized edge states in a semi-infinite system. We proposed to use a two-dimensional system of Rydberg-dressed atoms for experimental realization. The dressing gives rise to an effective interaction that vanishes on short interatomic distances, allowing us to neglect the atoms' spatial arrangement and treat it as a continuum. Besides the practical benefit, the continuum model is homogeneous, similarly to the quantum Hall effect. Our work leads to the open question of whether topologically non-trivial many-body states could be realized --- complementary to lattice-based fractional Chern insulators --- if we combine our continuum model with strong interactions. Because of the hard-core constraint of excitations and inevitable van der Waals interactions, strong interactions are naturally present in experimental realizations.

\begin{acknowledgments}
	We thank S. Hofferberth and K. Jachymski for interesting discussions.
	This work received funding from the European Union's Horizon 2020 research and innovation program under  the ERC consolidator grant SIRPOL (Grant No. 681208)
	as well as the French-German collaboration for joint projects in NLE Sciences funded by the Deutsche Forschungsgemeinschaft (DFG) and the Agence National de la 
	Recherche (ANR, project RYBOTIN). 
	P.B., acknowledge support by ARL CDQI, AFOSR, ARO MURI, DoE ASCR Quantum Testbed Pathfinder program (award No.\ DE-SC0019040), U.S.\ Department of Energy Award No.\ DE-SC0019449, DoE ASCR Accelerated Research in Quantum Computing program (award No.\ DE-SC0020312), NSF PFCQC program, and NSF PFC at JQI. 
\end{acknowledgments}

\bibliography{manuscript}
	
\ifsupp
\def\thesection{\Roman{section}}
\setcounter{secnumdepth}{2}
\widetext
\pagebreak
\ExecuteMetaData[supplement]{document}
\fi	
	
\end{document}


\renewcommand{\bibnumfmt}[1]{[S#1]}
\renewcommand{\citenumfont}[1]{S#1}

\title {Topological bands in the continuum using Rydberg~states}

\author{Sebastian~Weber}
\thanks{These authors contributed equally.}
\affiliation{Institute for Theoretical Physics III and Center for Integrated Quantum Science and Technology, University of Stuttgart, 70550 Stuttgart, Germany}
\author{Przemyslaw~Bienias}
\thanks{These authors contributed equally.}
\affiliation{Joint Quantum Institute and Joint Center for Quantum Information and Computer Science, NIST/University of Maryland, College Park, MD, 20742, USA}
\author{Hans~Peter~B\"{u}chler}
\affiliation{Institute for Theoretical Physics III and Center for Integrated Quantum Science and Technology, University of Stuttgart, 70550 Stuttgart, Germany}
	
\author{}


\makeatletter
\begingroup
\ltx@footnote@pop
\def\@mpfn{mpfootnote}%
\def\thempfn{\thempfootnote}%
\c@mpfootnote\z@
\let\@makefnmark\frontmatter@makefnmark
\begingroup
\frontmatter@title@above
\frontmatter@title@format
Supplemental Material for ``\@title"
\unskip
\phantomsection\expandafter\@argswap@val\expandafter{\@title}{\addcontentsline{toc}{title}}%
\@ifx{\@title@aux\@title@aux@cleared}{}{%
	\expandafter\frontmatter@footnote\expandafter{\@title@aux}%
}%
\par
\frontmatter@title@below
\endgroup
\groupauthors@sw{%
	\frontmatter@author@produce@group
}{%
	\frontmatter@author@produce@script
}%

\par
\frontmatter@finalspace
\endgroup

\makeatother

\setcounter{equation}{0}
\setcounter{figure}{0}
\setcounter{table}{0}
    
\renewcommand{\theequation}{S\arabic{equation}}
\renewcommand{\thefigure}{S\arabic{figure}}
	
\section{Simplified %
	model illustrating the cut-off potential} %
To illustrate the interplay of
blockade phenomena and dipolar exchange interaction, we present a simplified toy model. We consider two ground states $\ket{g},\ket{e_-}$ dressed with two Rydberg states $\ket{s},\ket{p_-}$.
\begin{figure}[H]
\centering
	\includegraphics[width= 0.30\textwidth]{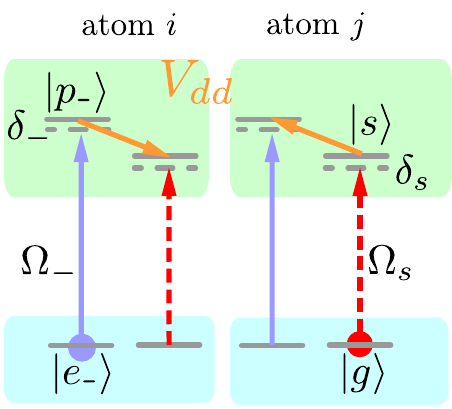}
	\includegraphics[width= 0.39\textwidth]{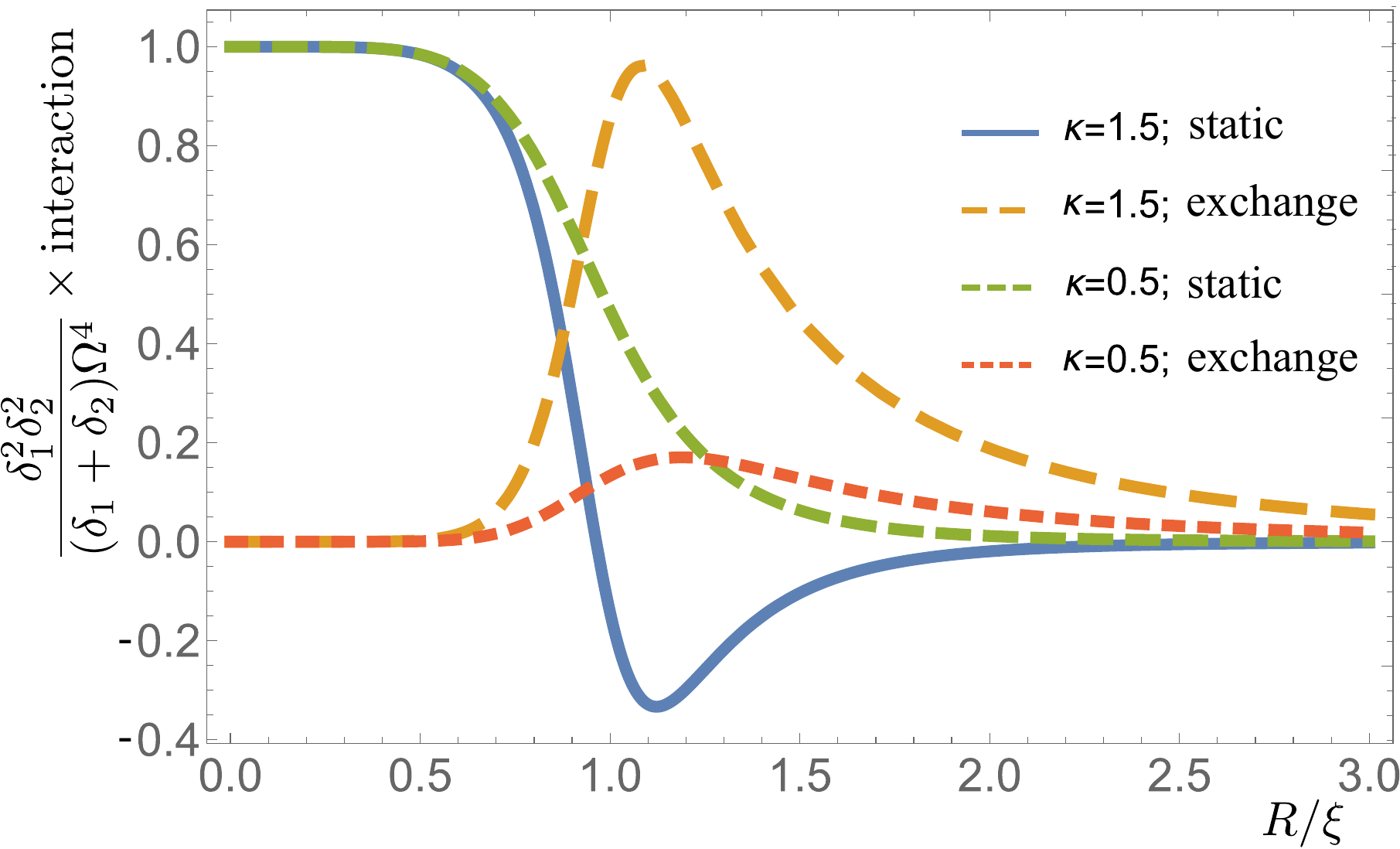}
	\caption{
		(a) The setup: Two identical atoms $i$ and $j$  are dressed with Rydberg states $\ket{p_-}$ and $\ket{s}$.
		(b) Effective interaction in fourth order between  dressed ground states $\ket{\tilde{g}}$ and $\ket{\tilde{e}_-}$ in the units characteristic for this problem (details in the text).
	}
	\label{fig:scheme}
\end{figure}

The single atom Hamiltonian $H_A$ and the laser drive Hamiltonian $\hat{h}$ read
\beqa
H^{(i)} 		&=&\d_s\ketbra{s}{s}_i+\d_-\ketbra{p_-}{p_-}_i\;,\nn\\
\HpertOp^{(i)}	&=& \OmEff %
\ketbra{g}{s}_i+\Om_-
\ketbra{e_-}{p_-}_i+\text{H.c.}
\eeqa
Dipolar interaction between Rydberg states is given by
\beqa
\Hdip&=&W_{ij} e^{-2i\phi_{ij}}\ketbra{s}{p_-}_i\ketbra{p_-}{s}_j\nn\\
&+&V_{ij}\ketbra{s}{s}_i\ketbra{p_-}{p_-}_j
+ h.c.\;,
\eeqa
where $W_{ij} =C_3/R^3$, $V_{ij} =C_6/R^6$, $R=|\br_i-\br_j|$. The total Hamiltonian reads
\beqa
\mathcal{H}=\sum_i H^{(i)}+\HpertOp^{(i)}+\sum_{i>j}V^{(i,j)}
\eeqa

We are interested in the interaction between weakly dressed ground states, which are denoted by $\ket{\tilde{g}}$ and $\ket{\tilde{e}_-}$.
In the limit of weak laser fields, we obtain effective interactions between ground states by treating the laser couplings \HpertOp as a perturbation~\cite{VanBijnen2015,VanVleck1929}. The static and exchange interactions between dressed ground states take the form
\beqa
\VdiagT&=&
\om_c\frac{\OmEff^2}{\d_s^2} \frac{\Om_-^2}{\d_{-}^2}
\frac{\Vdiag \left(\omC+\Vdiag\right)-\Vex^2}{\left(\Vdiag+\omC\right){}^2-\Vex^2}\;, %
\\
\VexT&=&
\om_c\frac{\OmEff^2}{\d_s^2} \frac{\Om_-^2}{\d_{-}^2}
\frac{\Vex \omC}{\left(\Vdiag+\omC\right){}^2-\Vex^2}\;,
\eeqa
where $\om_c=\d_-+\d_s$. We express the interactions using $\xi=(C_6/\omega_c)^{1/6}$, 
$\kappa=C_3\xi^3/C_6$ and $x=R/\xi$, giving rise to
\beqa
\VdiagT(R)&=&
- \om_c\frac{\OmEff^2}{\d_s^2} \frac{\Om_-^2}{\d_{-}^2}
\frac{\left(\kappa ^2-1\right) x^6-1}{\left(x^6+1\right)^2-\kappa ^2 x^6}\;,\\
\VexT(R)&=&
\om_c\frac{\OmEff^2}{\d_s^2} \frac{\Om_-^2}{\d_{-}^2}
\frac{ \kappa  x^9 }{\left(x^6+1\right)^2-\kappa ^2 x^6}\;.
\label{exchange}
\eeqa

Intuitively, the off-diagonal interactions   \VexT~can be thought of as 
microwave-assisted %
multi-photon processes leading to the hopping of the excitations --- which for the toy model involves four optical photons and an exchange of two microwave photons.

We see that in order to have resonance-free effective potentials, $\omC<0$ and $\kappa$ should be small enough so that for any separations $\left(x^6+1\right)^2-\kappa ^2 x^6>0$, and hence $\kappa<2$. The exchange potential height is 
\beqa
\max[\tilde{W}_{ij}] \approx \kappa\om_c {\OmEff^2}{\Om_-^2}/{(\d_{-}^2{\d_s^2} )}
\eeqa
for $\kappa\ll 1$.

We see that in the limit of small $\kappa$ we can approximate Eq.~\eqref{exchange} by Eq.~(2) in the main text, which motivated the use of this simplified formula for the illustration of the cut-off potential physics. Note that the schemes involving EIT dressing rather than off-resonant lead to quantitatively the same results; however, more lengthy expressions.


\bibliography{manuscript}